\documentstyle[11pt,aas2pp4]{article}

\def\mathmode#1{\ifmmode{#1}\else{$#1$}\fi}

\newcommand\et{{et al.}}

\newcommand\teff{\mathmode{{T}_{\rm eff}}}

\newcommand\etal{{et~al.}} %Current US (ApJ) standard no italics
\newcommand\lsun{\mathmode{L_{\odot}}}

\newcommand\mbfive{\mathmode{{m(162)}}}
\newcommand\bfivev{\mathmode{(162-V)}}
\def\ngc#1{{\rm NGC\thinspace{#1}}}
\newcommand\nred{NGC\thinspace362}

\def\msun{\mathmode{\,{M}_\odot}}
\def\flun{\mathmode{ {\rm erg\,s}^{-1}\,{\rm cm}^{-2}\,{\rm
\AA}^{-1}}}
\def\rhalf{\mathmode{{\rm R}_{0.5}}}

\slugcomment{Submitted to the Astrophysical Journal Letters}

\lefthead{Dorman \et}
\righthead{Hot Stars in NGC 362}

\begin{document}

\title{UIT Detection of Hot Stars in the Globular Cluster NGC 362 }

\author{Ben Dorman\altaffilmark{1,2}, Ronak Y. Shah\altaffilmark{3},
Robert W. O'Connell\altaffilmark{3}, Wayne B.  Landsman\altaffilmark{4}}
\author{Robert T. Rood\altaffilmark{3}, Ralph C. Bohlin\altaffilmark{5}, 
Susan G. Neff\altaffilmark{1}, Morton S. Roberts\altaffilmark{6},}
\author{Andrew M. Smith\altaffilmark{1}, and Theodore P.
Stecher\altaffilmark{1}}

\altaffiltext{1}{Laboratory for Astronomy \& Solar Physics, 
Code 681, NASA/GSFC,	Greenbelt MD 20771}
\altaffiltext{2}{NAS/NRC Resident Research Associate, NASA/GSFC} 
\altaffiltext{3}{Astronomy Dept, University of Virginia,
	P.O.Box 3818, Charlottesville, VA 22903-0818}
\altaffiltext{4}{Hughes/STX Corporation, 
Code 681, NASA/GSFC,	Greenbelt MD 20771}
\altaffiltext{5}{Space Telescope Science Institute, 3700 San Martin Drive,
Baltimore, MD 21218}
\altaffiltext{6}{National Radio Astronomy Observatory, 
Charlottesville, VA 22903}

% The abstract environment prints out the receipt and acceptance dates
% if they are relevant for the journal style.  For the aasms style, they
% will print out as horizontal rules for the editorial staff to type
% on, so long as the author does not include \received and \accepted
% commands.  This should not be done, since \received and \accepted dates
% are not known to the author.

\begin{abstract}

\vskip0.2in

We used the Ultraviolet Imaging Telescope during the March 1995
Astro-2 mission to obtain a deep far-UV image of the globular cluster
\nred, which was formerly thought to have an almost entirely red
horizontal branch (HB).
84  hot ($\teff > 8500\,$K) stars were detected
within a radius of 8{\farcm}25 of the cluster center.  Of these, 43
have FUV magnitudes consistent with HB stars
in \nred\ and at least 34 are cluster members.
The number of cluster members is made uncertain by  background
contamination from blue stars in the Small Magellanic Cloud (SMC).
There are six candidate supra-HB stars which have probably
evolved from the HB.  We discuss the implications of these results for
the production of hot blue stars in stellar populations.  

\end{abstract}

\keywords{globular clusters: general---
globular clusters---individual (NGC 362)---stars: evolution---
stars: horizontal-branch---ultraviolet---stars}

\twocolumn

\section{Introduction}

The temperature structure of the horizontal branches of globular
clusters is controlled by the distribution of envelope masses of stars
as they reach the zero-age-horizontal branch (\cite{rtr70}; \cite{do92})
after losing mass on the red-giant branch (RGB).  The mass and
metallicity of the envelope govern the temperature of stars on the
ZAHB and their later evolution.  For $M < 1 \msun,$ smaller envelopes
produce hotter stars, other things remaining constant.  All clusters
show a spread in HB color which implies a finite range in envelope
mass.  But clusters have very different HB temperature distributions,
some with entirely hot stars and others with entirely cool (i.e. $B-V
\gtrsim 0.2$) stars (reviewed in \cite{zinn93}).  This variety of HB
morphologies can, in principle, be produced by (i) differences in
abundances, which affect the opacity of the envelope or
equilibrium in the hydrogen burning shell; (ii) differences
in the masses of stars on the RGB due to age, and 
thus in the HB masses assuming mass loss
is similar among clusters, or (iii) variations in net mass loss, 
 amounting to $\sim 0.2$--0.4\msun, during RGB evolution.

Most of the variation in HB morphology among Galactic globular clusters
correlates with differences in the mean metal abundance, which has
therefore been called the ``first parameter''.  However, observed
differences in HB's at a given metal abundance imply that it is
necessary to invoke at least one more factor, i.e., a ``second
parameter.''  Many argue (e.g., \cite{ldz94}) that the
``global'' second parameter is age, i.e. that in general the underlying
cause behind the bluer HBs is a greater age because of (ii) above. 
The identification of age as the second parameter is, however, highly
controversial (e.g., \cite{svb96}).

In this paper we report the detection on far-ultraviolet images of hot
HB stars in \nred, an important globular cluster well known for having
red HB morphology, and in a companion paper (\cite{oc47}) we report
similar detections in 47 Tuc.  \nred\ is often compared with
\ngc{288}, which has similar abundance but a blue HB; this pair is one
of the most important for second parameter studies (\cite{bo89}; 
\cite{gn-n362}; \cite{sd90}; \cite{vbs90}, \cite{det91}).  Ground-based
color-magnitude diagrams (CMDs) show that \nred\ has an almost  entirely
red HB. \cite{ldz94} using data from \cite{har82} gave its 
HB type index $\rm  (B-R)/(B + V + R) = -0.87,$ based on
$\rm B:V:R = 3:4:77$ from photographic data in $R > 1\farcm57$.
\cite{har82} detected 3 stars with colors blueward of the instability strip;
however radial velocities indicate that the membership of two of these
is doubtful (R.~C.~Peterson 1995, private communication). 

 The detection of blue HB stars in clusters with red HB's is important
because it implies that RGB mass loss is either bimodal---perhaps
the result of more than one mass loss process---or that the range of
RGB mass loss is larger than had previously been thought.

\section{Observations}

We observed \nred\ with the Ultraviolet Imaging Telescope (UIT) and
obtained deep, wide-field images at 1600 \AA.  The UIT is a 38-cm
Ritchey-Chr\'etien telescope with a 40\arcmin\ diameter field of view
and solar-blind detectors for imaging at vacuum-UV wavelengths (see
\cite{uit} for details).  It flew as part of the {\it Astro-2\/} UV
observatory on the Space Shuttle {\it Endeavour\/} during 1995 March.
The data were recorded on IIa-O film, which was scanned with
microdensitometers at 20$\mu$m resolution at Goddard Space Flight
Center. The resulting density images were converted to intensity,
flat-fielded, and flux-calibrated using the batch data reduction
procedures developed for the purpose (\cite{hil96}).  The calibration
datasets used here are the  ``Flight 22'' versions.  The resulting
images have a plate scale of 1{\farcs}14/pixel.  Point sources on the
\nred\ image have FWHM $\sim$ 4\arcsec\ owing to jitter in the pointing
system aboard the Shuttle.

UIT  images of \nred\ were obtained with the far-UV $B5$ filter, which has
a peak wavelength $\lambda_0 = 1620 {\rm\:  \AA}$ with width $\Delta
\lambda = 230 {\rm\:  \AA}$ (\cite{uit}).  Here, we use the longest
exposure for the analysis (frame FUV2897; exposure time 808.5 s)
since it is nowhere saturated.  Astrometry was obtained using a 
combination of the
\cite{tuc92a} and the CCD images of Montgomery \& Janes (1994,
hereafter \cite{mj94}), using objects in common with the
HST Guide Star Catalog.  The far-UV image of cluster is shown in
Figure 1 (Plate X).

Photometry was obtained using an Interactive Data Language (IDL)
implementation of DAOPHOT I (\cite{daop}), which has been modified to
accommodate the noise characteristics of film.  Photometry was
performed both using aperture methods and point spread function (PSF)
methods, with results that agree within the PSF fitting/sky background
errors except in the few crowded regions.  Typical errors for the UIT
photometry are 0.15 mag, including uncertainty in the aperture
correction of 0.10 mag owing to a variable PSF.  We quote FUV
magnitudes on the monochromatic system, where $m_{\lambda}(\lambda) =
 -2.5 \log(F_{\lambda}) -21.1$, and $F_{\lambda}$ is in units of \flun.
We refer to $B5$ magnitudes
below as \mbfive\ and $B5-V$ colors as \bfivev.

Because of UIT's large field and the suppression of the cool main
sequence and RGB in the UV, the samples of UV-bright stars (generally
with $\teff > 8000 {\rm \, K}$) we find are complete, even in the
cluster core.  Due to the failure of UIT's mid-UV camera on the {\sl
Astro-2} mission, however, we require ground-based photometry to
estimate temperatures.  Kent Montgomery 
and Kenneth Janes have kindly provided us with CCD $BVI$ photometry of
\nred, a summary of which can be found in \cite{mj94}.
For  the common region (13 square arcmin) excluding the crowded
center ($\sim 30\arcsec$), we use the
\cite{mj94} data to construct UV-optical color-magnitude diagrams.

The cluster lies near the Small Magellanic Cloud (SMC)
in projection, and it
is necessary to consider contamination by background main sequence
stars in the Cloud.  As basic parameters we adopt the
following (\cite{sw86}, \cite{tkd95}, \cite{sgd93}):  $(m-M)_0 =
14.67$; $E(B-V) = 0.04$; and a half-light radius $\rhalf = 40 \arcsec.$
The cluster center is taken to be $\rm \alpha_{J2000} = 01^h\, 03^m \,
14{\fs}3,$ $\rm \delta_{J2000} = -70^\circ \, 50\arcmin \, 54\arcsec.$
In computing the FUV apparent distance modulus, we have adopted the
Galactic UV reddening law of \cite{ccm89} according to which
$A(B5)/E(B-V) = 8.06$.  Estimates for the tidal radius of the cluster
range from $9{\farcm}85$ (\cite{tuc92b}) to $14{\farcm}85$
(\cite{tkd95}).

\section{Results}

We have identified a total of 84 stars ($3\sigma$ detections) on the UIT
image in a area 16{\farcm}5\ square centered on the cluster.  14 of the
detections lie in a knot that coincides with the optical center and is
$ < 14\arcsec$ in radius (see Fig.~\ref{fig:n362}).  There are 17
detections in common with the bluest stars found on recent HST WFPC2
images of the cluster center in the F218W, F439W filters (C.  Sosin
1996, private communication), including this central clump.  Of these
detections, 49 match blue (i.e. $B-V < 0.2$) stars in the MJ94 list within
$r = 8{\farcm}25$. 24 of these lie above the theoretical ZAHB or
within $1\,\sigma$ of it ($\mbfive < 15.4$ for stars
at the blue end of the HR diagram). We have chosen not to
explore potential identifications with red MJ94 stars; further work may
reveal these to be hot binary companions to cool stars.

The stars detected span a large range in temperature.
Fig~\ref{fig:cmd362} shows the UV-optical CMD we derive from the
matched stars within a radius of 8{\farcm}25\ of the center.  The
colors we measure range from $-3.72 < \bfivev < 0.54,$ corresponding to
effective temperatures of
$23000 >  \teff > 8500 {\, \rm K}.$ We plot the ZAHB
and evolutionary sequences from \cite{dro93} adopting ${\rm [Fe/H]}
=-1.48$, ${\rm [O/Fe]} = 0.6$. To the left is the histogram of UIT
detections that do not have blue optical counterparts in the
\cite{mj94} data.  Of the 3 blue stars mentioned by \cite{har82}, the
only  radial velocity member (H1328) is clearly detected on the UIT
image with $\bfivev = 0.35.$

The sources that lie near or above the plotted ZAHB in
(Fig.~\ref{fig:cmd362}) are thus possible members of an \nred\ blue HB
sequence.    The 35 stars for which we do not have colors include stars
too crowded for identification in the $V$-band image (including all of
the stars of the central 14\arcsec),  out of the \cite{mj94} field, or
very hot objects that are intrinsically faint in $V$ but not in
\mbfive.  19 of these detections have UV magnitudes which are brighter
than 1-$\sigma$ below the ZAHB level at the blue extreme of the ZAHB.
Adding these to the 24 stars plotted on the CMD, we estimate there are
43  stars in \nred\ with UV magnitudes consistent with membership of
this blue HB sequence.  For the majority of these stars, either their
location near the center of the cluster ($R \lesssim \rhalf$), or their
location on the HR diagram with $\bfivev > -2,$ strongly suggests
membership (a total of 32 stars).  In addition, there are six stars
(four on the CMD and two in the histogram) with $\mbfive < 14.$  A
magnitude or more brighter than the theoretical ZAHB, these are
candidates for post-HB, evolved objects.

The field of \nred\ is contaminated by young main sequence stars in the
SMC.  \cite{bo87} conducted deep photometry of fields to the north-east
of the cluster and found a well-defined young (300 Myr) main sequence
extending continuously to $(V,B-V) = (19.5,-0.2),$
and four blue stars that are about 1
mag brighter\footnote{According to the field locations
given in \protect\cite{bo87}, these stars are detected by UIT with $\mbfive$
between 14.4 and 15.9.}.  There is   a pronounced gradient in the blue
star counts across the field. To complicate estimates
of the background counts the image suffers from vignetting effects in
the part of the field most strongly
influenced by the SMC.  In order to measure the
contamination, we have counted stars in the largest circle centered on
the cluster that does not suffer from the vignetting problem,
$12{\farcm}04$ in radius.  The number of UIT sources with $R > 3
\arcmin$ lying below a line running south-east to north-west is nearly
three times the number of detections above it: there are almost no
stars in the NE quadrant of the image.  There are a total of 179 stars
in this wider field, 113 of which have $\mbfive > 15.4,$ i.e. fainter
than the expected brightness of cluster EHB stars. There appears to be
an excess of these (16) in the area occupied by the central 3\arcmin of
the cluster. Possibly there are some EHB stars in the cluster that are
underluminous (cf. \cite{w94}).  However it is probable
that the irregularity as well as the steep gradient
of background counts across this field makes
simple estimates of the contamination unreliable.

For the 6 stars lying $> 1$ magnitude above the HB we estimate that  at
most one of the post-HB candidates should be an early-B SMC background
object.  This is based on the number of mid-B SMC stars apparently
present and assuming a Salpeter IMF. There is also a small probability
of contamination by foreground stars, but it is unlikely that there
will be more than one in this field. Altogether, assuming at least 32
stars close to the ZAHB are members plus at least 4 supra-HB stars, we
conclude that there are 36--43 blue HB stars in the cluster.

If all of the detections with $14.2 < \mbfive < 15.2$ without optical
matches are EHB stars, the supra-HB stars potentially have 22 unevolved
counterparts (i.e. they are within the range of core He burning
evolution off the ZAHB).  This gives a ratio of 4:1, which is
consistent with the theoretical expectation, based on stellar lifetimes
(\cite{dro93}).  However, if a significant fraction of these stars are
in fact cooler, the situation may more resemble that of M3
(\cite{buz92}). This cluster has two hot, bright objects ($\teff \sim
30,000\, {\rm K},$ $\log L/\lsun \sim 2$) with no obvious progenitors.
In this case the bright stars may arise through binary evolution.

We used the \cite{mj94} CMD to compute the relative numbers of blue and
red HB stars. For simplicity the numbers we quote below are upper
bounds taking the contamination as zero.  There are 292 red HB stars
with $R > 30\arcsec$, inside which the radial histogram shows obvious
incompleteness effects, and 29 blue HB stars. 
The latest version of the Sawyer Hogg catalogue (C. Clement 1996, private
communication) lists 7 confirmed RR Lyrae stars in the cluster, 
giving $B-R/(B + V + R) \sim -0.78.$

Finally, note that the exposure has
insufficient depth to reach the blue straggler sequence, which would be
2 mag fainter than the ZAHB at $\bfivev = 0.$

The integrated apparent \mbfive\ magnitude of \nred\ is found to be
$10.22\pm 0.11$ to a radius of 8{\farcm}25; this includes all
contributions from hot stars which are too faint to be individually
resolved.  This is only an upper bound because of the uncertainty due
to contamination.
Its integrated visual  magnitude is $V = 6.40,$ \cite{har82}---see
also \cite{mo85}---implying $\bfivev_0 = 3.62.$ OAO-2 (\cite{wc80}) 
measured $m_\lambda(2460)_0 = 8.66
\pm 0.08.$ The resulting far-to-mid UV color of $1.23 \pm 0.14$ places
\nred\ at the red end of the cluster sequence in Fig.~3 of
\cite{dor95}.  This color is consistent with a small blue HB
population.  A simple estimate
of the expected value of the integrated FUV/optical color may be
derived---assuming the ratio of blue to red HB stars including the
central 30\arcsec\ is not very different from that computed from the
outer region---as follows.  Using the Fuel Consumption Theorem, and
the flux ratio of late to early stages we obtain $F_V(HB + AGB)/F_V(GB
+ MS) = 0.25.$ Then, assuming N(BlueHB)/N(Red HB)$ = 0.1,$ the ratio
FUV/V is $\sim 1/40$, reliable to about $\pm 50\%,$ or $\bfivev_0 \sim 4 \pm 0.4.$ 

\section{Discussion}

Although the detection of blue HB stars in a cluster with a
predominantly red HB does not conform to canonical expectations, no
new physics is necessary to explain these observations. The novelty in
the observations reported here lies in the fact that a globular cluster
often cited as having an almost purely red HB has a significant blue
HB component with few objects at intermediate temperatures. The
simplest interpretation is that RGB mass loss in this cluster is
elevated for a minority of the RGB stars, producing the lower mass
blue HB component.  The derived integrated UV/optical color of
\nred\ is comparable with a number of ellipticals and spiral bulges
(\cite{dor95}), so that even the small number of stars
we have detected here can have some bearing on the ``UV upturn''
phenomenon in galaxy populations.  

Nearly all of the synthetic modeling of HB stellar mass distributions
uses unimodal mass distributions.  However, the \nred\ HB
morphology observed here is inconsistent with the hypothesis that RGB
mass loss is dominated by a process with only
normally-distributed stochastic scatter superposed.  Either there is
more than one mass loss process affecting HB morphology, or
alternatively a single process may lead to a highly non-Gaussian
distribution (for a discussion see
\cite{ddro96}).  Star-to-star variation in stellar rotation rates is one
potential cause for the scatter on the HB.  Mixing during RGB evolution
(perhaps also induced by rotation) can increase the He content of the
H-burning shell regions of ZAHB stars, which also produces hot HB stars
without increased mass loss (\cite{sw96}).  However, the observational
record is unclear: the few studies of HB rotation that have been
completed, albeit for lower temperature stars (i.e. $\log\teff < 4.1;$
\cite{prc95}), do not show a trend of rotational velocities with
temperature within a given cluster. 

\nred\ has been regarded by many authors as a cluster with an
HB morphology that is ``too red'' for its metallicity, although
estimates for $\rm [Fe/H]$ do range as high as $\sim -1$ (\cite{det91}).
As such it has frequently been
taken as a test case for the notion that HB morphology can be used as
an age indicator.  Currently, the age determination of \nred\ based on
main sequence and subgiant branch morphology is controversial and
unlikely to be settled immediately (\cite{svb96}).  However, detection
of blue HB stars in this cluster is significant whether or not a young
age is confirmed.
For whatever the age of \nred\, its CMD is inconsistent with a
simple unimodal HB mass distribution. Of course this is not the first
example of bimodality or other deviations from the usual assumptions
used in HB synthesis; others are the famous example of NGC 2808
(\cite{fer90}; \cite{so97}), NGC 1851 (\cite{wa92})
and the gaps in the horizontal branches of M15 and NGC
288 (\cite{cro88}). The increasing evidence that more than one process
affects the structure of globular cluster HB's suggests that
ages for clusters derived solely from HB morphologies are 
unreliable (\cite{dor95}).

If the turnoff fit does imply that the cluster is
$\sim 2$--3\,Gyr younger than the majority of the Galactic globular
cluster system, then stellar systems with measurable UV flux can be
produced at a relatively young age ($\sim 11$--12\,Gyr).  This would
contradict the hypothesis that far-UV radiation is always an indicator
of great age for extragalactic stellar populations (e.g. \cite{pl97};
\cite{bcf94}), which relies heavily on the assumptions of unimodality
and predictability of the HB mass distribution with age and metallicity.

\acknowledgments

We are very grateful to Kent Montgomery and Kenneth Janes for providing
their CCD photometry tables to us for both clusters in advance of
publication and to Craig Sosin and George Djorgovski for pre-release
data from HST. B.D. would like to thank both Robert S. and Robert J.
Hill for many useful conversations about the data reduction.  We thank
Bob Cornett for providing the SMC theoretical main sequence, and
Joel Parker for the ground-based image used in Figure 1.
We would also like to acknowledge use of WWW resources placed
online by J.-M. Perel  and W. E. Harris. We would like 
to thank the referee, Michael Bolte, for helpful comments. Parts of
this research have been supported by NASA grants NAG5-700 and
NAGW-4106 to the University of Virginia and NASA RTOP 188-41-51-03.

\onecolumn
%%% REFS  

\clearpage

\begin{figure}
\vskip 1.0truein
\centerline{\bf bitmap file  ngc0362.gif goes here}
\vfil
%\plotone{ngc0362.ps}
\caption{\protect\label{fig:n362} Far-ultraviolet (1600 \AA) image of
\nred, extracted from UIT frame FUV2789 and showing a square of side
$16{\farcm}50$ centered on the cluster.  The solid circle indicates a 3
\arcmin\ radius, while the dashed circle shows the half-light radius at
40\arcsec.  The lower inset show a close-up of the central knot of
UV-bright sources, while the upper inset shows a corresponding
$V$-band image kindly provided by Joel Parker from data taken at CTIO
in 1995 December.  The dashed circles in the inset are the same radius
as in the main image.}
\end{figure}

\clearpage

\begin{figure}[t]
\vspace{-1.90in}
\plotfiddle{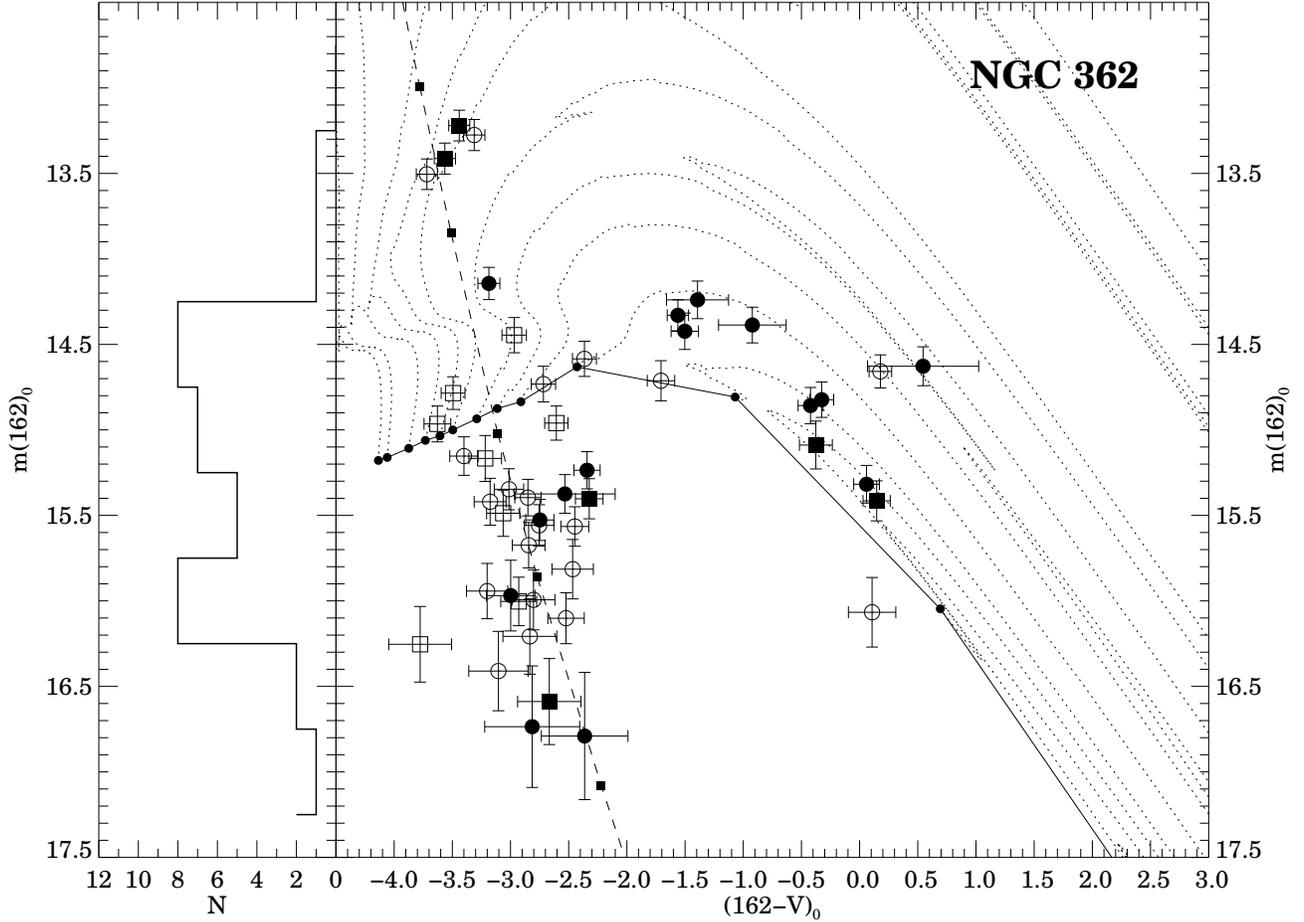}{\hsize}{90}{65}{65}{256}{0}
\caption{\protect\label{fig:cmd362} The color-magnitude diagram of NGC
362 constructed from our data and the $V$ band observations of 
\protect\cite{mj94}, which cover a 13\arcmin\ square field 
centered on the cluster.
The symbols indicate spatial location on the image:
filled symbols are stars in the central 3\arcmin, open
symbols outside of this radius. Circles denote sources below the SE to
NW line referred to in the text (where SMC background contamination is
largest), while squares show detections above this line. We assume
$E(B-V)= 0.04$ and $(m-M)_0 = 14.67.$ The solid line with the small
filled circles shows the ZAHB, with the post-HB evolutionary tracks
shown by dotted lines.  The evolutionary sequences are from
\protect\cite{dro93}, with [Fe/H] $= -1.48,$ [O/Fe] $= 0.60.$ The
``mean age'' SMC main sequence at $(m-M)_0 = 18.9$ is shown by filled
squares joined by a dashed line, from models by 
\protect\cite{char} 
and composition parameters for the SMC given by \protect\cite{wes90}.
The synthetic magnitudes are derived from
 \protect\cite{k92} model stellar fluxes.  To the left of the figure
is the histogram of stars without optical identifications but within
the \protect\cite{mj94} field.   }
\end{figure}

\end{document}